# ARTICLE INFORMATION

### Article title

EweAcT: Ewe behaviour aligned to accelerometer data for activity monitoring in extensive grazing systems.

### Authors

Lucile Riaboff*[,1], Ny Aina Andriamampandry[1], Jean-François Bompa[1], Mathias Aletru[1], Christian Durand[2], Sébastien Douls[2], Gaëtan Bonnafe[2], Morgane Costes-Thiré[1], Guillaume Delosières[1], Jean-Marc Mongrelet[1], Enzo Niro[1], Némuel Tadi[1], Séverine Deretz[2], Sara Parisot[2], Margot Lamarque[2], Dominique Hazard[1], Emilie Cobo[1].

### Affiliations

[1] GenPhySE, Université de Toulouse, INRAE, ENVT, 31326, Castanet-Tolosan, France

[2] UE0321 INRAE La Fage, 12250 Saint Jean et Saint Paul, France

### Corresponding author's email address and Twitter handle

lucile.riaboff@inrae.fr

### Keywords

Accelerometer collars, Sheep, Behaviour, Annotation, Extensive grazing systems.

### Abstract

Monitoring livestock behaviour under extensive conditions would provide valuable insights to assess animal adaption to environmental perturbations in agroecological systems (e.g., heat waves, parasitism, predator attacks). Animal behaviour can be monitored using accelerometer data collected from neck-collars combined with artificial intelligence models. However, large amounts of accelerometer data aligned with annotated behaviours are necessary to develop accurate models of behaviour prediction. In particular, developing reliable models for extensive systems requires data collected across a wide range of representative conditions. The dataset includes 79 hours of tri-axial accelerometer data aligned with behaviours manually annotated from video recordings for 120 Romane ewes born between 2021 and 2024. The ewes were derived from two divergent genetic lines after three and four generations of selection started 10 years ago: low and high social attractiveness, noted S- and S+, and low and high tolerance towards humans, noted H- and H+. They were reared under the extensive system applied to the Experimental Unit of La Fage (UEF, INRAE, Saint-Jean-et-saint Paul, Aveyron) where 250 sheep were reared exclusively outdoors on 280 hectares of rangeland in southern France. First batch of data was collected on March, June and July 2024 at the UEF under a range of extensive conditions, including sloping pastures and heat-wave periods. Ewes were equipped with accelerometer neck-collars specifically designed for young sheep on pasture. They were grouped on experimental paddocks for 4 to 8 hours and provided with fresh grass and *ad libitum* access to water. The animals were simultaneously video-recorded using an elevated CCTV camera. Behaviour annotation was carried out using Behavioral Observation Research

Interactive Software focusing on the main behaviours on pasture: Grazing, Ruminating, Resting, Moving, and “Other”, grouping all remaining activities. Annotations and corresponding accelerometer sequences were aligned using Python language, based on a time synchronization procedure. A second batch of data was acquired on November 2025 to supplement the dataset with the moving activity. For that purpose, ewes were equipped with the accelerometer collars and moved on tracks from the housing area to the pastures, corresponding to an approximately 10 minute-walk. The start and end times of the moves for each ewe were used to align the corresponding accelerometer data with the moving activity. These data were then merged with the dataset from the first batch. The resulting dataset is ready to use for applying artificial intelligence models to classify the 5 main behaviours of sheep under extensive grazing systems from accelerometer data.

## SPECIFICATIONS TABLE

| | |
|---|---|
| **Subject** | Computer Sciences |
| **Specific subject area** | Sheep behaviour monitoring from accelerometer data under extensive systems. |
| **Type of data** | Table; Raw. |
| **Data collection** | Romane ewes from 2 divergent genetic lines for social traits were equipped at the La Fage Experimental Unit (INRAE) with a tri-axial accelerometer (AX3 Axivity) inserted into a neck-collar [1]. Ewes were placed in paddocks and recorded with CCTV cameras fixed high up. The main behaviours were manually annotated from videos using Behavioral Observation Research Interactive Software. Timestamp synchronization was applied to align annotations with accelerometer data. Ewes were also moved from housing area to pastures for recording additional moving activity. Dataset includes 79 hours of accelerometer data from 120 ewes aligned to behaviour. No normalization was applied. |
| **Data source location** | Institution: UEF, INRAE. Experimental Unit of La Fage.<br>City/Town/Region: Saint-Jean-et-Saint-Paul, Aveyron<br>Country: France<br>Latitude/Longitude: 43.9182319, 3.0944099 |
| **Data accessibility** | Repository name: data.gouv.fr<br>Data identification number: https://doi.org/10.57745/AM2AJI<br>Direct URL to data:<br>https://entrepot.recherche.data.gouv.fr/dataset.xhtml?persistentId=doi:10.57745/AM2AJI |
| **Related research article** | None |

## VALUE OF THE DATA

- The EweAcT dataset is specifically designed to monitor sheep behaviour in extensive systems from accelerometer data, encompassing a wide range of conditions, such as sloping paddocks, different seasons and varying animal thermal states. To the authors' knowledge, this is the first dataset specifically designed for livestock behaviour monitoring in this agroecological system.
- EweAcT includes 79 hours of annotated data from 120 Romane ewes from 2 divergent genetic lines selected for social traits (high/low sociability toward conspecifics; high/low docility toward humans). The large amount of data covering a wide diversity of conditions will contribute to the development of robust and high-performing Artificial Intelligence (AI) models to classify sheep behaviour on pasture from accelerometer data.
- EweAcT will enable AI models development for continuous monitoring of sheep behaviour in extensive conditions over long periods, providing new insights into sheep behaviour reared exclusively outdoor in an extensive system (e.g., baseline behaviour in the absence of disturbance and behavioural responses to environmental perturbations). Such insights can further be used to explore several levers for animal adaption to agroecological systems (e.g., social skill transfer, genetic selection).
- Continuous and individual behaviour monitoring will enable the study of relationships between behaviour, health, and welfare in extensive conditions.
- This dataset combines accelerometer time series with annotations used as a gold standard. coreThe time series and their alignment with the annotations were carefully inspected to ensure data quality and reliability. In this regard, EweAcT is ready to use for the development of new deep learning models for the field of time-series classification.

## BACKGROUND

Compliance with the core principles of agroecology promotes extensive systems in which animals are reared outdoors. However, such agroecological systems expose animal to more variable environment including various disturbances (meteorological hazards, resource scarcity, predation, etc.) that can affect animal health and welfare [2]. Maintaining good animal welfare is however essential to bridge the gap between farming and society, to decrease health issues and to maintain production. As farm animals are social species, animals' behavioural adaptability to their environment (e.g., anti-predator strategies) depends mainly on their social skills [3]. Animal docility should also help to reduce acute stress during human interventions [4]. In that regard, Hazard et al. [5] selected divergent genetic lines for social and human responsiveness. Measuring how young sheep adjust their behaviour in response to environmental disturbances to maintain their health and welfare, and whether genetic selection for higher sociability or greater docility leads to better adaptation, requires continuous behaviour monitoring over long periods in extensive systems. Behaviour monitoring can be achieved using accelerometer data, collected from neck collars and analyzed with AI models [6]. However, large amounts of accelerometer data aligned with annotated behaviours collected under conditions representative of extensive systems are required for model development.

## DATA DESCRIPTION

The EweAcT dataset is a table saved as a PARQUET file (*EweAcT.parquet*; 62.5 Mo) containing 9 columns and 7,108,285 rows. A description of each column and format is provided in Table 1. Each row represents an accelerometer reading along three dimensions (columns name: *AccX, AccY*, and *AccZ*) collected from one animal at a specific timestamp (sampling rate: 25 Hz) and aligned with a behaviour (*Behaviour*). The focus was on the main behaviours of sheep on pasture: Grazing, Resting, Ruminating, Moving, and Other, that merges the other activities such as grooming, social interactions, or standing/lying transitions. Definitions of the behaviours are provided in Table 2. Complementary information (e.g., posture, abnormal behaviour, type of social interaction) are also provided in the dataset when appropriate (*Modifier_1*, *Modifier_2*, *Modifier_3*). Comments done during the annotation process were also provided (*Comment*). In total, the dataset comprises 79 hours of accelerometer time series collected from 120 ewes. Figure 1 shows the number of accelerometer readings (or rows of the dataset) for each behaviour and the number of ewes associated.

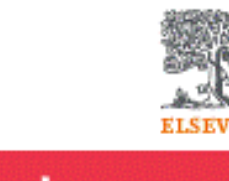

Table 1. Name, description and format for each column of the EweAcT

| Column name | Description | Format |
|---|---|---|
| AnimalID | Animal ID associated with the accelerometer reading. | Integer |
| DateTime | Timestamp of the accelerometer reading with sampling rate = 25 Hz | Timestamp (Y-M-D H:M:S.ffffff) |
| AccX | Accelerometer reading for the X-axis | Float |
| AccY | Accelerometer reading for the Y-axis | Float |
| AccZ | Accelerometer reading for the Z-axis | Float |
| Behaviour | Annotated behaviour associated with the accelerometer reading, *i.e.*, Grazing, Ruminating, Resting, Moving or Other (see Table 2). | String |
| Modifier_1 | First level of additional information to the behaviour observed (see Table 2); NaN was used when no additional information was provided. | String |
| Modifier_2 | Second level of additional information for the behaviours Resting and Other (see Table 2); NaN was used when no additional information was provided. | String |
| Modifier_3 | Third level of additional information for the behaviour Other (see Table 2); NaN was used when no additional information was provided. | String |
| Comment | Comment associated with the accelerometer reading, done during the annotation process (e.g., “Looks agitated”). | String |

Table 2. Definition of the five behaviours in the *Behaviour* column with their modifiers

| Behaviour | Definition | Modifiers[1] |
|---|---|---|
| Grazing | Biting grass with the head down and chewing for less than 5 sec | Exploring: Moving with the head down for a few steps while searching for new grass |
| Ruminating | Chewing and regurgitating bolus- chewing interruptions < 10 sec | Lying: Ruminating while lying; Standing: Ruminating while standing |
| Resting | Lying or standing still, without ruminating | Lying: Resting while lying **>** Curled up/On ground/Straightened: Position of the head; Standing: Resting while standing |
| Moving | Moving from one location to another without grazing | Active Walking: Walking head up in the experimental paddock; Moving along the track: Walking or trotting on the track while moving the ewe from housing to pasture; Running: Running in the paddock |
| Other | All the activities other than grazing, ruminating, resting and moving | Bleating: Bleating mouth opened; Drinking: Drinking in the water trough; Grooming: Self-grooming without social interaction **>** Type of grooming behaviour: Licking / Scratching / Shaking / Stretching / Other **>>** Body part the ewe is grooming: Body / Forelegs / Head / Limbs / Rear limbs; Urinating; Defecating; Panting: Deep, heavy breaths with mouth open or closed **>** Posture while panting: Lying / Standing; Jumping; Social interaction: Interaction with at least one congener **>** Type of social interaction: Fighting / Grooming / Sniffing / Other; Transition: Transition between lying and standing **>** Change of posture: Lying down / Standing up |

[1] Main modifiers are underlined and followed by secondary and tertiary modifiers, indicated using the “**>**”, “**>>**” symbols for secondary and tertiary, respectively.

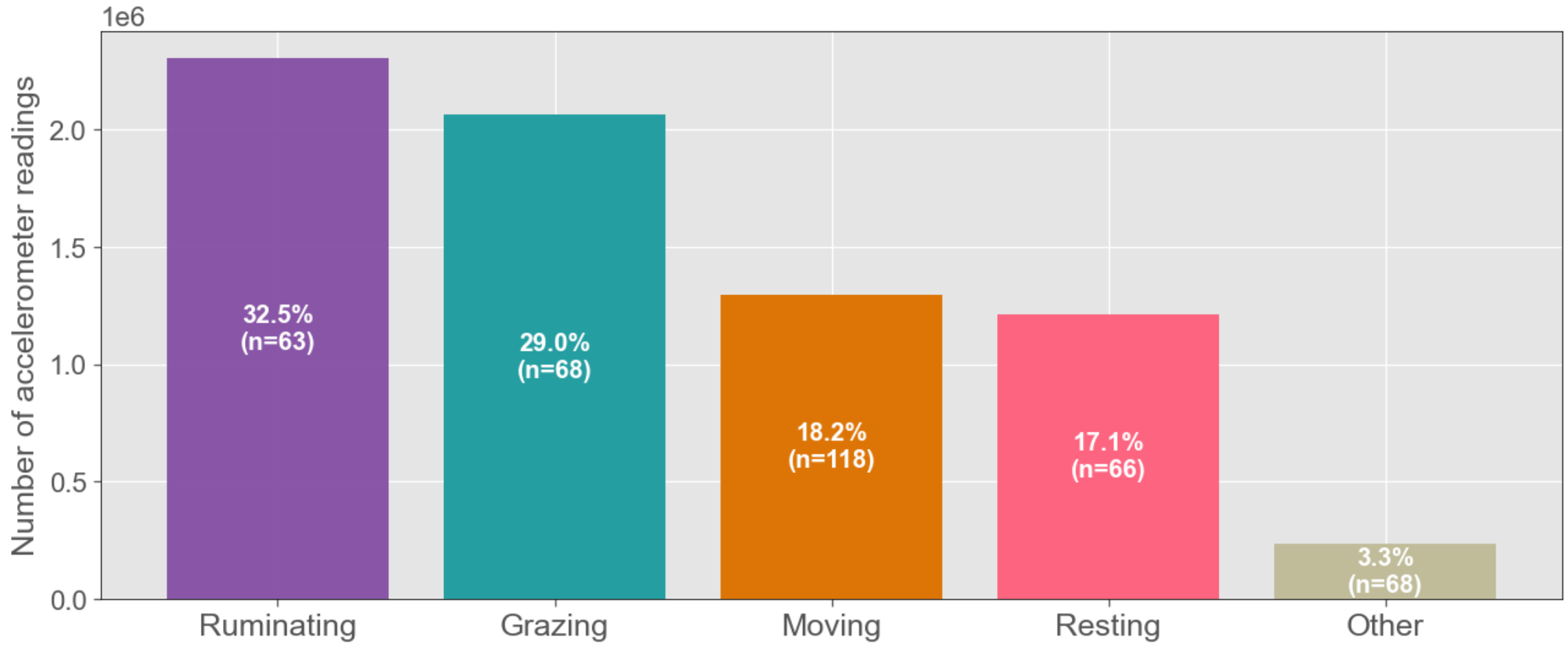

Figure 1. Number of accelerometer readings (x1e6) in the dataset for each behaviour. Percentages of specific reading among total readings are shown within each bar; number of ewes (noted n) among the 120 ewes annotated in the dataset for each behaviour is indicated in brackets. We refer to Table 2 for a definition of each behaviour.

A metadata table saved as an Excel file (*metadata_EweAcT.xlsx*; 14 KB) is provided in addition to the dataset. This table contains the key characteristics associated with EweAcT. It includes 13 columns and 153 rows. Each row contains information for an individual ewe (column name: *AnimalID*), including the ID of the AX3 sensor (*AX3*) worn during the period in which the ewe was video-recorded or directly observed (*time_start_observation*; *time_stop_observation*), as well as the exact times (UTC) at which the accelerometer sensor was started and stopped (*time_start_accelerometer*; *time_stop_accelerometer*). The table includes animal genetics (*genetics*), year of birth (*year_of_birth*), the month of the year when the animal was observed (*month_of_year*), the location where it was observed (*location*), whether the experimental paddock was sloped (*slope*), whether the animal showed signs of heat stress during the annotation process (*thermal_status*), and the average outdoor temperature during the period when the animal was filmed (*T_avg*). It should be noted that the same animals could be observed during different periods (e.g., June 2024 and November 2025) and under different conditions (e.g., both in flat and sloped paddocks). In such cases, there are as many rows in the metadata for a given animal as there are different experimental conditions. A description of each column, including criteria for each characteristic, is provided in Table 3. The percentage of annotated accelerometer data associated with each characteristic is shown in Figure 2.

The Jupyter Notebook (*AX3BehaviourAlignment. ipynb*), developed in Python (v3.11.7) to create the EweAcT dataset, has also been added to the repository. Sample files, including an extract from the AX3 accelerometer sensors (*93942_34057_2024-06-11.csv*), a file exported from the BORIS software (*34057_2024-06-11_090929_Cam1_LR.csv*), and a sample of the dataset under construction (*2026-05-27_EweAct_sample.parquet*), have also been included. These files allow the notebook to be executed successfully and provide readers with a step-by-step illustration of how the EweAcT dataset was constructed.

Table 3. Name and description of the columns in the metadata with criteria within the key characteristics (genetics, year of birth, month of year, slope, thermal status, average outdoor temperature)

| Column name | Description (format) | Criteria |
|---|---|---|
| AnimalID | Animal ID for which behaviours were annotated (format: integer) | x |
| AX3 | AX3 accelerometer id (format: integer) | |
| time_start_observation | Start date and time when the animal was video-recorded or directly observed (format: d/m/Y H:M:S) | x |
| time_end_observation | End date and time when the animal was video-recorded or directly observed (format: d/m/Y H:M:S) | x |
| time_start_accelerometer | Start date and time when the accelerometer worn by the animal was started (format: d/m/Y H:M:S) | x |
| time_stop_accelerometer | Stop date and time when the accelerometer worn by the animal was started (format: d/m/Y H:M:S) | x |
| genetics | Divergent genetic line for social traits of the ewe (format: string) | H+: High docility toward humans; H-: Low docility toward humans; S+: High sociability to conspecifics; S-: Low sociability to conspecifics |
| year_of_birth | Year of birth (format: integer) | 2021, 2022, 2023 or 2024 |
| month_of_year | Month during which the ewe was filmed (format: string) | March, June, July, November |
| location | Location where the ewe was filmed or observed (format: string) | Paddock; Track |
| slope | Presence or absence of slope in the experimental paddock where the ewe was filmed (format: string) | Flat; Sloped |
| thermal_status | Presence or absence of signs of heat stress observed during the behaviour annotation process (format: string) | Thermal comfort: no panting observed; Heat stress: panting with the mouth open or closed observed when the outdoor temperature was > 20 °C. |
| T_avg | Average temperature (°C) when the ewe was video-recorded or directly observed, calculated from the hourly temperatures recorded at the weather station (median: 12.5°C [min: 6.4°C; max: 25.1°C]), and then turned into a range of temperature (format: string) | < 10°C: average outdoor temperature below 10°C<br>[10°C, 20°C]: average outdoor temperature between 10°C and 20°C<br>> 20°C: average outdoor temperature above 20°C. |

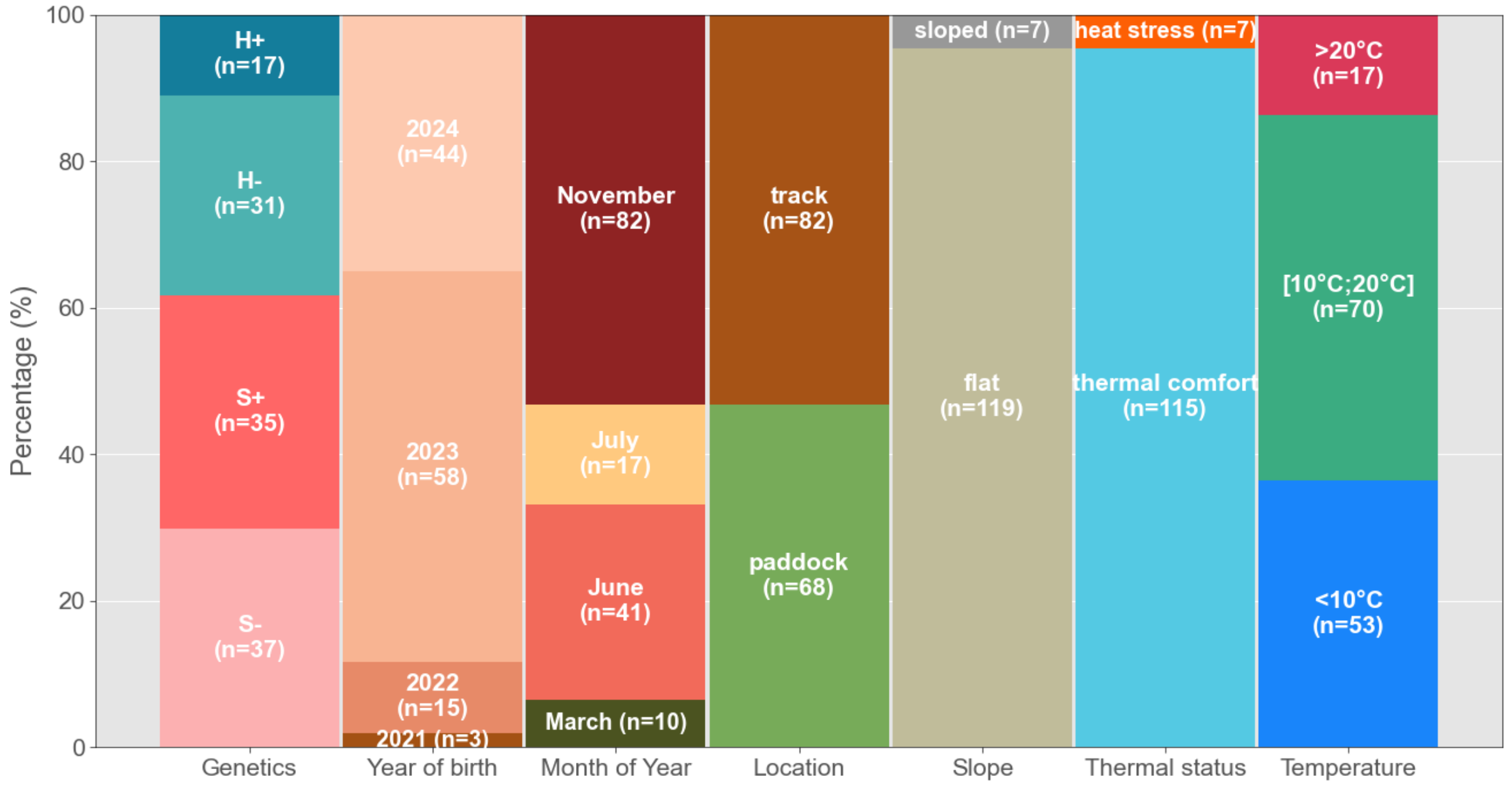

Figure 2. Percentage of annotated accelerometer *data* associated with each key characteristic, as defined in Table 3. Criteria for each characteristic are written in the corresponding segment; number of ewes (noted n) observed for each criterion is indicated in brackets. It should be noted that month, location, slope, thermal status and temperature are not exclusive characteristics, as the same ewe could be observed in different experimental conditions. We refer to Table 3 for a description of the characteristics and criteria.

# EXPERIMENTAL DESIGN, MATERIALS AND METHODS

1. Materials

*Accelerometer sensors*

Eighty tri-axial accelerometers (AX3, Axivity Ltd, Newcastle, UK) were used to record accelerometer data. The dataloggers weigh 11 g and measure of 23 × 32.5 × 7.6 mm (see Figure 3). They were configured with a sampling rate of 25 Hz (i.e., 25 readings per second) and a sensitivity of ±4 g, resulting in a battery life of over 30 days. The OmGui interface developed by the manufacturer was used for dataloggers configuration. Battery charging was performed by connecting each accelerometer to a computer *via* a USB-C cable, taking approximately 1h30 min to reach a full charge. Data were retrieved from the accelerometers and downloaded in CWA format using the OmGui interface. To facilitate and speed up the process, a system was developed within INRAE (CATI SICGPAE, France) allowing 10 accelerometers to be inserted into 10-compartment cylinders connected to eight hubs, each equipped with 10 USB ports (see Figure 4) [1]. The CWA files were then saved to a hard drive.

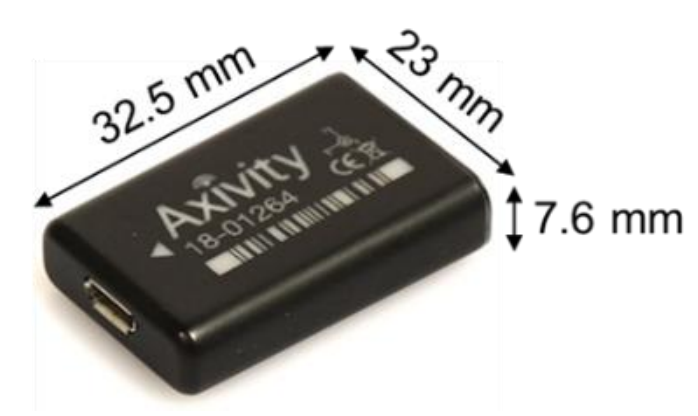

Figure 3. 3-axis accelerometer
(AX3, Axivity Ltd, Newcastle, UK)

10-room cylinder holding 10 accelerometers
Computer with OmGui interface
Hub with 10 USB ports
USB-C Cable
Hard drive

Figure 4. Homemade equipment designed for accelerometer battery charging, configuration, and data downloading [1].

*Accelerometer neck-collars*

Accelerometer neck-collars [1] were specifically developed for the experiment within INRAE (CATI SICGPAE, France) after several trials performed in experimental farms Langlade (https://doi.org/10.17180/ftvh-x393) and La Fage (https://doi.org/10.15454/1.548325523466425E12). Each accelerometer was encapsulated into a locker. The locker was inserted into a hole located on the right side of the dorsal part of the neck collar and secured with a detachable rivet (see Figure 5). The dorsal part (weight: 80 g; dimensions: 48 × 23 × 12 mm) is a rigid, rounded structure made of glycol-modified polyethylene terephthalate material, designed to avoid any discomfort or injury to the ewes. Four colors (green, orange, blue and purple) were used for the dorsal part of the collar to distinguish animals depending on their genetic lines. The neck collar was fitted around the ewe's neck using the adjustable strap (see Figure 6). Once fitted around the ewe's neck, the accelerometer's X-axis aligns with the head-to-tail direction, the Y-axis with the lateral direction, and the Z-axis with the top-to-bottom direction. To prevent collar rotation, a counterbalance weight (100 g) was fixed at the bottom of the ventral part. This design ensures that the accelerometer is positioned exactly the same way from one ewe to another with no spurious movements, while providing waterproof and mechanical protection to the sensor.

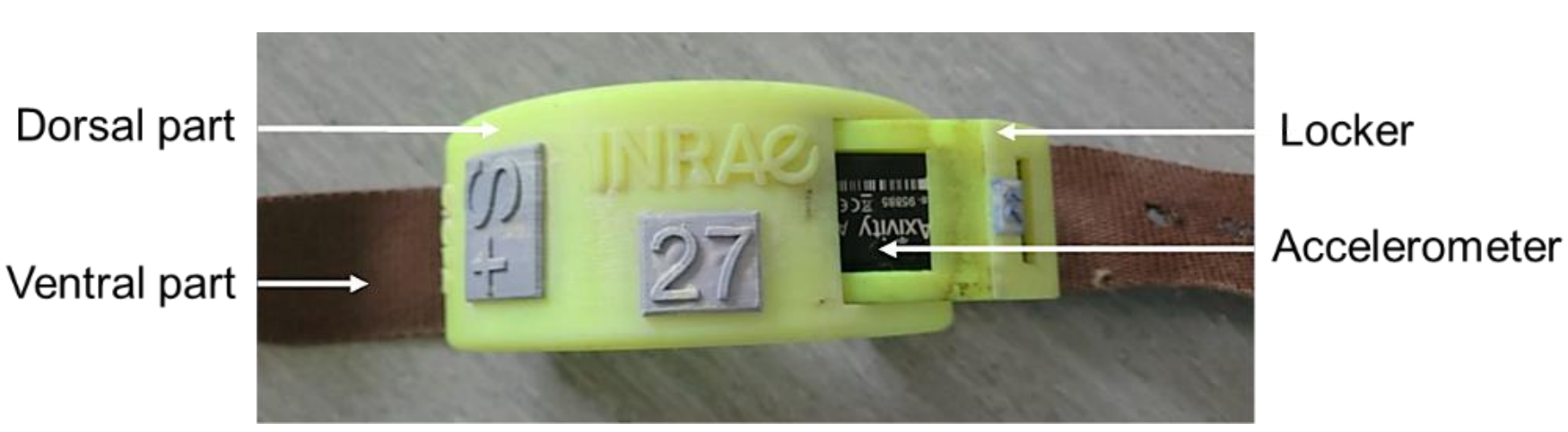

Figure 5. Neck-collar designed for the experiment [1]. The accelerometer is encapsulated into the locker and inserted into the dorsal part of the collar. The ventral part allows fine adjustment of the collar around the ewe's neck.

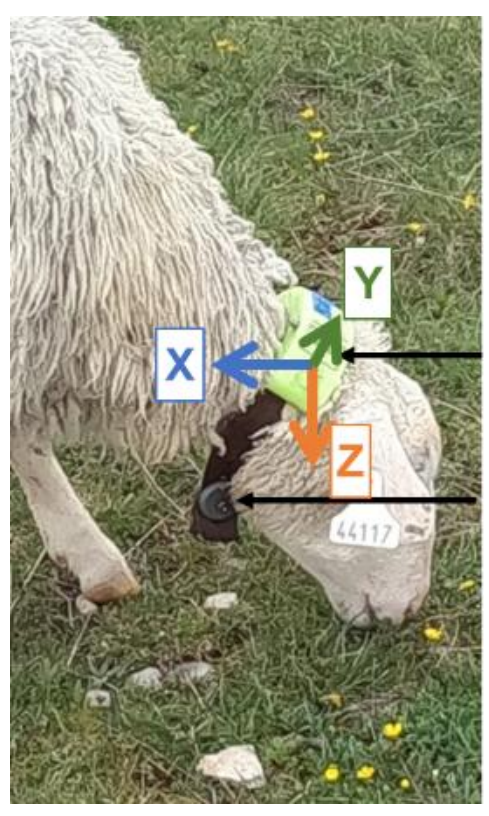

Figure 6. Accelerometer neck-collar adjusted around the ewe's neck. The counterweight (100 g) prevents collar rotation. X, Y, and Z correspond to the 3 axes of the accelerometer sensor: X corresponds to the head-to-tail axis, Y to the lateral axis, and Z to the top-bottom direction.

*Cameras*

The video footage recording system was developed within INRAE (CATI SICGPAE, France) [1]. Four CCTV cameras (DAHUA) were used for video recording (25 frames/second). The cameras were powered either by a mobile battery or from a 220-volt mains supply through a switch connected *via* a network cable (see Figure *7*). The switch was also connected to a Raspberry Pi (Pi3 Model B) through ethernet cable. Cameras were started and stopped from the Raspberry Pi interface; the corresponding times aligned to the Raspberry Pi UTC clock were also recorded in the media name. A TP-Link wireless access point installed inside the switch allowed a computer to connect to the switch *via* a Wi-Fi network so that video recordings could be accessed using the SmartPSS software. Videos footages were transmitted to the Raspberry Pi through the switch and saved to the Raspberry Pi hard drive. Video footage was recorded in H.264 format at a resolution of 1920 × 1080 pixels and was converted to MP4 format.

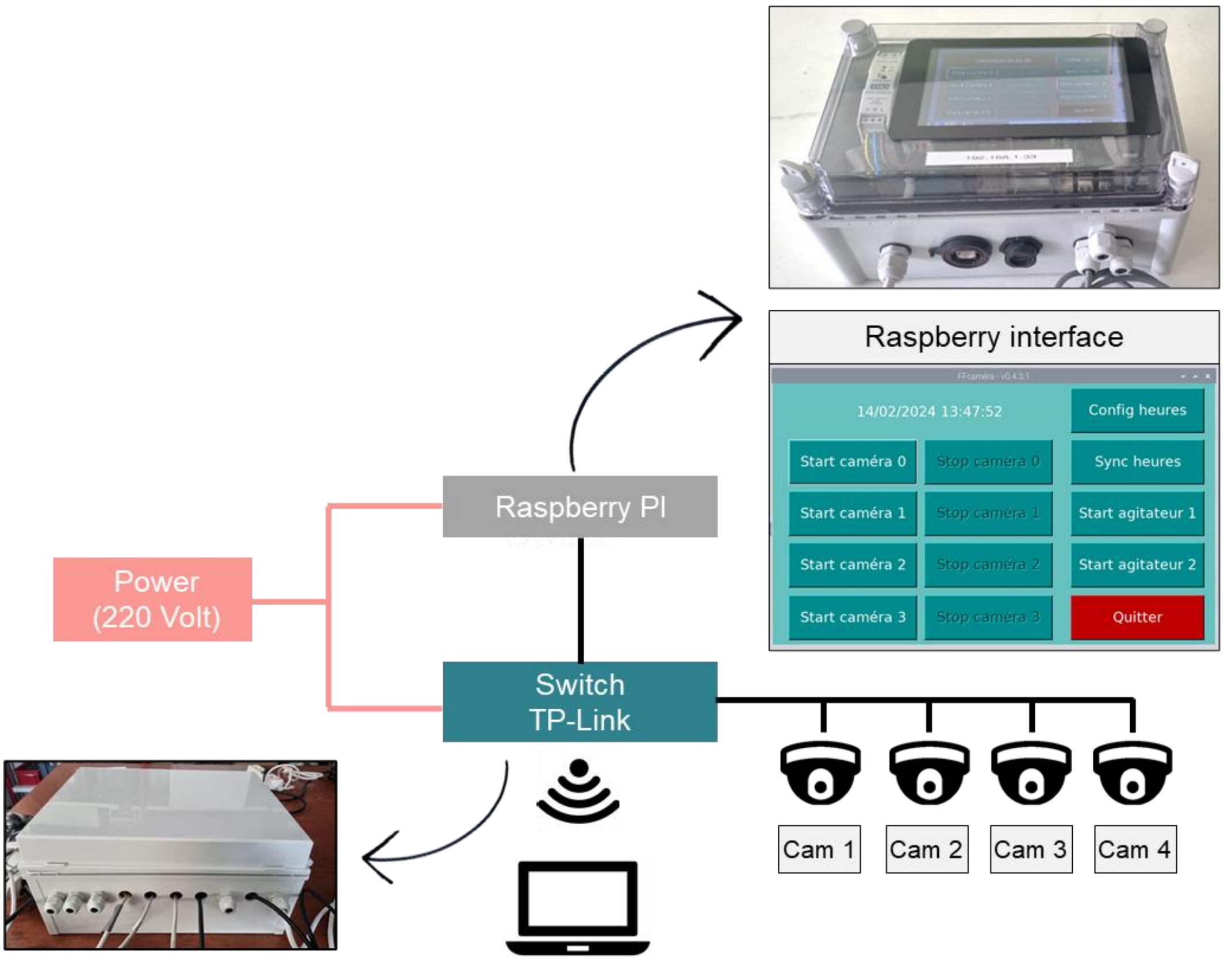

Figure 7. Video recording system developed in and used to videotape the animals [1]. The four cameras were powered *via* a 220-volt supply through a switch, which was also connected to a Raspberry Pi. The cameras were started and stopped through the Raspberry Pi interface. A Wi-Fi network was provided by a TP-Link wireless access point integrated into the switch.

2. Experiment

*Location*

Experiment was carried out at the Experimental Unit of La Fage (INRAE, Saint-Jean-et-Saint-Paul, Aveyron, https://doi.org/10.15454/1.548325523466425E12) on 25 March 2024, from 10 to 13 June 2024, from 23 to 26 July 2024 and on 06 and 13 November 2025. Ethical approval was obtained from the Ethics Committee in Science and Animal Health (No. 115) of the Veterinary School of Toulouse (Toulouse, France) under authorization ID SSA_2024_003V2.

*Animals*

Ewe lambs from divergent genetic selections for social traits (H+: high docility toward human; H-: low docility toward human; S+: high sociability to conspecifics; S-: low sociability to conspecifics) born between 2021 and 2024 were used for the trial. Ewes were reared exclusively outdoor with few inputs according to the extensive system applied at the Experimental Unit of La Fage.

*Data collection*

*[Batch 1]*

The first batch of data was collected on 25 March 2024, from 10 to 13 June 2024 and from 23 to 26 July 2024. The aim of this batch was to film the ewes equipped with accelerometer neck-collars while exhibiting the full range of behaviours on pasture. For that purpose, 148 ewes born in 2023 (n = 68) or 2024 (n = 80) from each of the divergent genetic lines (H-: n = 37; H+: n = 36; S-: n = 38; S+: n = 37) were equipped with the accelerometer collar. On each trial day, accelerometer sensors were configurated (sampling rate: 25 Hz; sensitivity: ±4 g), time synchronized, and attached to the necks within the following hour. Animal IDs and corresponding collar IDs were carefully recorded. The ewes were then housed in groups of five in one of four experimental paddocks (12 × 12 m) for 4 to 8 hours, provided with fresh grass and *ad libitum* access to water. Each group of five animals was filmed with a camera mounted 4 m above the ground in each paddock. The paddock size, camera height and angle were previously optimized to allow the ewes to express natural activities on pasture while enabling accurate behaviour annotation from the videos without blind spots. Each ewe in a paddock was marked on its back with a unique color. The back color, collar color, and paddock ID for each ewe were recorded to facilitate subsequent identification in the video recordings. From 23 to 26 July, one of the four paddocks was sloped to collect accelerometer data under typical conditions of the extensive system. It should be noted that temperatures higher than 20°C were recorded during this period, resulting in signs of thermal discomfort and heat stress in the ewes. The experiment time was thereby shifted from early morning (6:00 am) to early afternoon (1:00 pm), with shadow availability in the paddocks lasting until approximately 11:00 am, to balance animal welfare considerations with the need to record data under an extreme thermal condition often encountered in extensive systems. Accelerometer data and videos were retrieved at the end of each trial day.

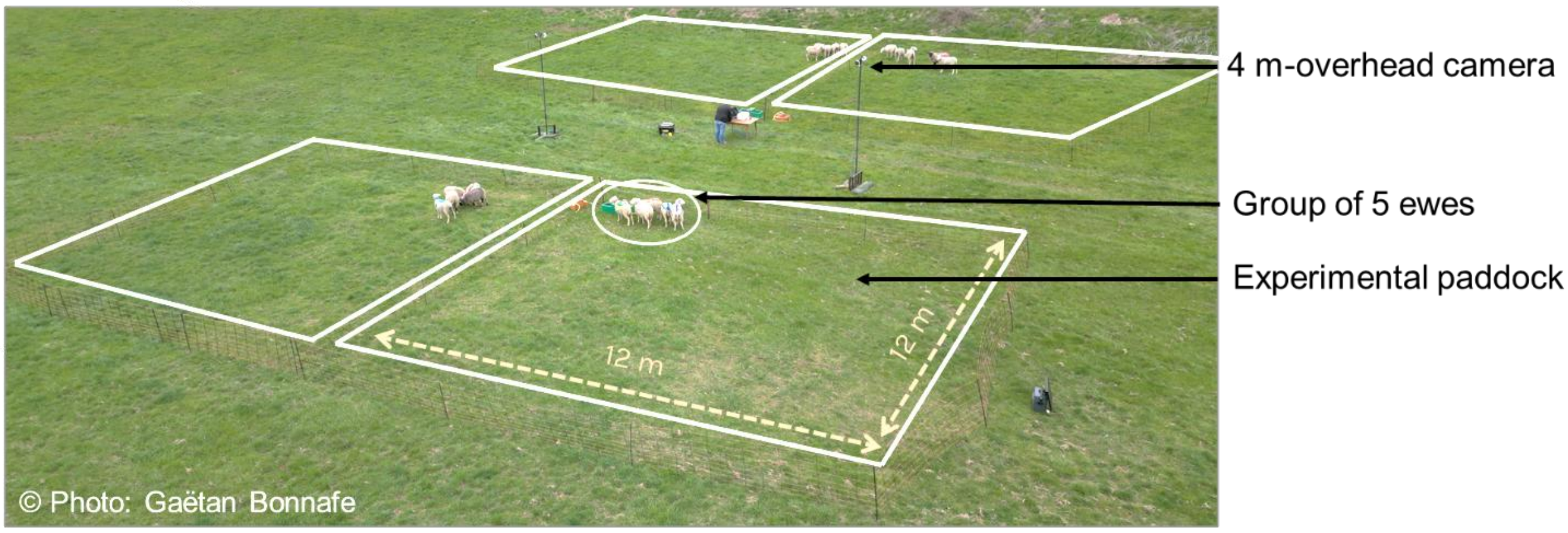

Figure 8. Experimental design for collecting data for the full range of behaviours [Batch 1]. Ewes were moved into 12 × 12 m experimental paddocks in groups of five and filmed for 4 to 8 hours using an overhead camera.

*[Batch 2]*

The second batch of data was collected on 06 and 13 November 2025. The aim of the second batch was to collect accelerometer data while the ewes were actively moving, as this activity class was underrepresented in the first batch of data. 82 ewes born in 2021 (n = 3), 2022 (n = 15), 2023 (n = 31) and 2024 (n = 33) from three of the divergent genetic lines (H-: n = 24; S-: n = 29; S+: n = 29) were used for the second batch. Of these, 24 ewes born in 2023 and 6 ewes born in 2024 had also been part of batch 1. For this purpose, the ewes were fitted with the accelerometer neck-collar (sampling rate: 25 Hz, sensitivity: ±4g) and then moved for approximately 10 minutes according to their genetic line, in one of ten moving groups consisting of 7 to 20 animals. Animal IDs, corresponding collar IDs, moving group and the start and end times of each move, recorded using a smartphone clock, were noted for all animals. At least two people were involved in each moving group: an animal caretaker from the Experimental Unit led the ewes at the front using a barley bucket, while another person positioned behind the group encouraged the ewes to keep moving and prevented them from deviating from the track to graze (see Figure 9). The latter also filmed the ewes during the move with a GoPro camera (HERO) fixed to the chest. Any issues (e.g., ewes stopping, deviating from the path to graze) and the associated times were audio recorded. The ewes were required to move along different types of tracks (paved, grass, and dirt), thereby reflecting the diverse terrain that may be used by the ewes in extensive grazing system. Accelerometer data were retrieved on the 21 November 2025 during the next routine weighing of the ewes.

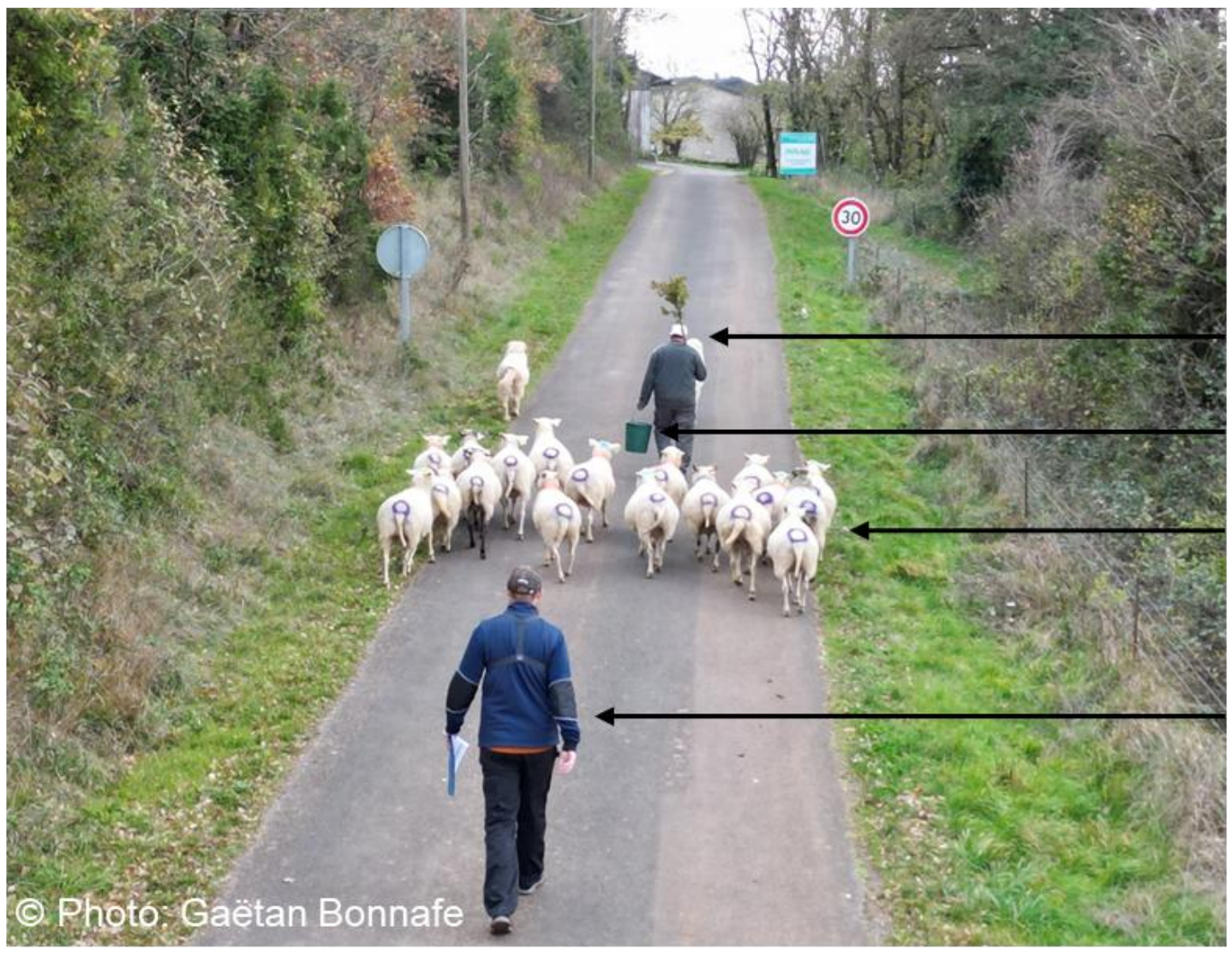

Figure 9. Group of ewes walking from housing to pasture to collect moving data [Batch 2]. At least one animal caretaker led the group from the front using a bucket of barley, while at least one experimenter followed behind to encourage the animals to walk.

3. Time synchronization

Time synchronization was applied to the first batch of data. This step aimed to align the timestamps of the accelerometer data and the timestamps of the video records to a common clock (UTC clock), ensuring that the annotated behaviours could be accurately matched to the corresponding accelerometer data. The approach used in this experiment relied on generating a distinct and specific artificial pattern in the accelerometer signal, for which the corresponding time was known and synchronized with the video clock. For this purpose, two cylinders, each holding 10 accelerometers, were designed to rotate for 10 seconds, resulting in a characteristic and repeatable pattern (see Figure 10). The cylinders were connected to the Raspberry Pi and activated *via* the interface. The exact time, accurate to the millisecond according to the Raspberry Pi's UTC clock, was automatically saved in a ".txt" format as soon as the cylinders were activated to the Raspberry Pi's hard drive. Since the cameras were connected to the Raspberry Pi (see Figure 7), the time recorded was on the same UTC clock used for the video recordings. The pattern in the accelerometer signal was automatically detected with the *signal.find_peaks* function applied to the Y-axis (parameters: height = 3.99; width = 17 * sampling rate) from the *scipy* package in Python language (V3.11.7). For each accelerometer sensor, the difference between the time in the accelerometer data associated with the detected pattern and the time recorded by the Raspberry Pi was calculated. The entire accelerometer time series was then corrected for this time offset to align to the Raspberry Pi UTC clock. Furthermore, it should be noted that a time drift may occur in the accelerometer timestamps, resulting after 4 to 8 hours in a time offset of several seconds between the accelerometer data and the video recordings. For this purpose, the cylinders were activated both before attaching the accelerometers to the neck collars and after removing the sensors. The difference between accelerometer time and Raspberry Pi time at the second cylinder activation was used to linearly adjust each timestamp in the accelerometer time series. A private function was implemented in Python (V3.11.7) to (i) detect the pattern in the accelerometer signal, (ii) correct the timestamps of the entire time series using the UTC time associated with the pattern at the first activation, and (iii) correct for time drift at each timestamp using the UTC time associated with the pattern at the second activation. This function, named *time_sync*, is available in the Jupyter Notebook (*AX3BehaviourAlignement.ipynb*) provided in the data repository. It can be run using the AX3 accelerometer file included in the repository as an example (*93942_34057_2024-06-11.csv*) together with the metadata file (*metadata_EweAcT.xlsx*; see Table 3).

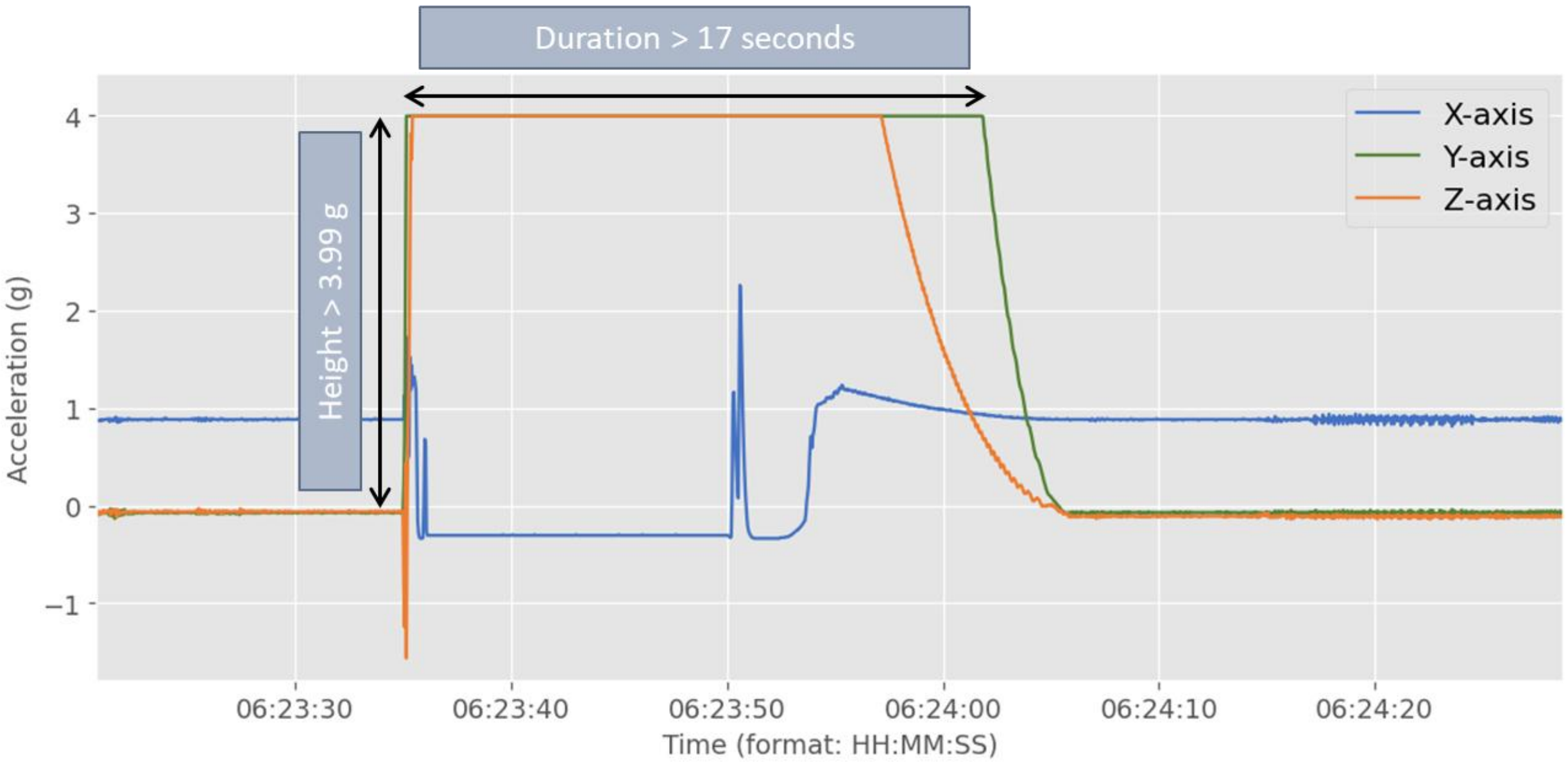

Figure 10. Pattern generated in the accelerometer signal when activating the cylinder. The X-axis, Y-axis and Z-axis are displayed in blue, green and orange, respectively. The Y-axis was used for aligning the accelerometer time to the UTC time of the Raspberry Pi. The parameters used in the signal.find_peaks function (height, duration) to detect the peaks automatically with Python language are shown in the Figure.

4. Behaviour annotations

For the first batch of data, behaviours were annotated from the videos records using the Behavioral Observation Research Interactive Software (BORIS) [7]. An ethogram based on the five main behaviours of sheep on pasture (Grazing, Ruminating, Resting, Moving, Other), including modifiers (see Table 2), was defined prior to the annotation process. Videos were imported into the BORIS software in MP4 format. Annotation focused on a single animal at a time, identified by its paint color, collar color, and experimental paddock ID. Each selected video was partially annotated by clicking on the observed behaviour in the BORIS interface, defined as a "state event." The start and end times relative to the beginning of the video for each annotated behaviour were automatically recorded (see Figure 11). Any abnormal behaviours were also recorded. In particular, panting (heavy breathing with mouth closed or open when the temperature exceeded 20°C in the absence of shade on the pasture) was systematically noted. Sections of the videos in which behaviour was difficult to identify, due to the ewe being too far from the camera, in a shadowed area, or the behaviour being hard to distinguish, were not annotated. As the annotation process progressed, greater focus was placed on underrepresented behaviours to help balance the dataset; thus, sections of the videos in which the behaviour of interest was exhibited were preferentially annotated. Once annotation of a given video was complete, the BORIS file was exported in CSV format, with the animal ID and the video's start time in UTC included in the file name to facilitate merging with accelerometer data.

Annotation process was carried out by two observers. To minimize observer bias, a training period was first conducted for each observer. Observers were allowed to discuss with each other during this period to clarify behaviour definitions when necessary. After training, the behaviours of two animals exhibiting a wide range of activities were both carefully annotated (total annotation duration: 1 hour) by each observer, based on the 5-behaviour ethogram including all modifiers (see Table 2). During this stage, observers performed the task individually without talking to each other. Cohen's Kappa [8] was then calculated at 0.5-second intervals to measure agreement between the observers throughout the annotation process. A Cohen's Kappa > 0.80 was obtained, indicating strong agreement; therefore, the annotated behaviours were considered reliable and independent of the observer. Overall, 70% of the video footage recorded during batch 1 was used for annotation: behaviours were manually annotated for 68 out of 148 ewes, corresponding to 71 hours of accelerometer data aligned to a behaviour.

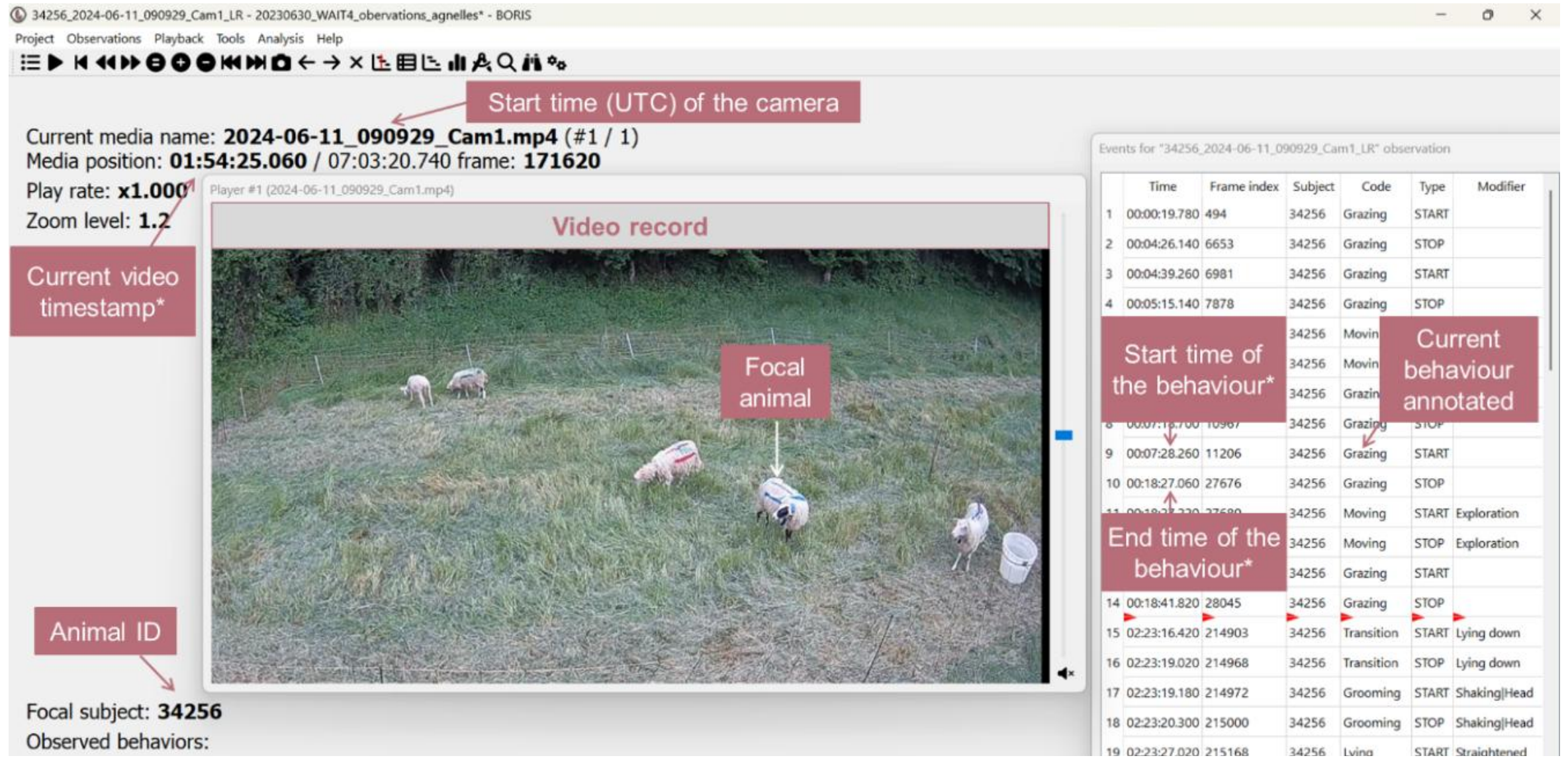

Figure 11. Boris interface for behaviour annotation of a focal animal from a video record.
Note: *The timestamps displayed are relative to the beginning time of the video.

5. Accelerometer data and annotated behaviour merging

The final step consisted of merging the behavioural annotations with the corresponding accelerometer data. For batch 1, accelerometer data were first aligned to the UTC clock of the Raspberry Pi and corrected for drift, as described in the “Time synchronization” section. Timestamps associated with each annotated behaviour, extracted from CSV files generated by BORIS, were converted to a datetime format based on the start time of the videos, which had been aligned to the Raspberry Pi UTC clock (see “Material” section). Each annotated accelerometer sequence was carefully inspected by a domain expert to find any inconsistencies between the accelerometer patterns and the annotated behaviours, which could arise from animal or paddock misidentification, or errors during the annotation process. The sources of error were identified for all sequences with inconsistencies, which were subsequently corrected before being added to the dataset. Any remaining time offsets of a few seconds between the start of the annotated behaviour and the corresponding accelerometer data (min: 0 s; max: 3.2 s; mean: 0.42 s) were also corrected. It should be noted that all portions of the accelerometer sequences that were not annotated were excluded from the dataset, resulting in breaks in the continuous sequences in EweAcT. A Python script (V3.11.7) was developed to facilitate the inspection of annotated accelerometer sequences, as well as their correction, integration into the dataset and tracking of the sequences merging. This script is provided in the Jupyter Notebook (*AX3BehaviourAlignement.ipynb*) available in the data repository. It can be run using the AX3 accelerometer file (*93942_34057_2024-06-11.csv*), the BORIS annotation file (*34057_2024-06-11_090929_Cam1_LR.csv*), and a sample of the dataset (*2026-05-27_EweAcT_sample.csv*) included in the repository as examples, together with the metadata file (*metadata_EweAcT.xlsx*; see Table 3). Newly annotated accelerometer sequences can then be progressively merged into the dataset, while an updated record of the sequences added is maintained in the text file *BorisAlignementTracking.txt*. Sequences of annotated accelerometer data

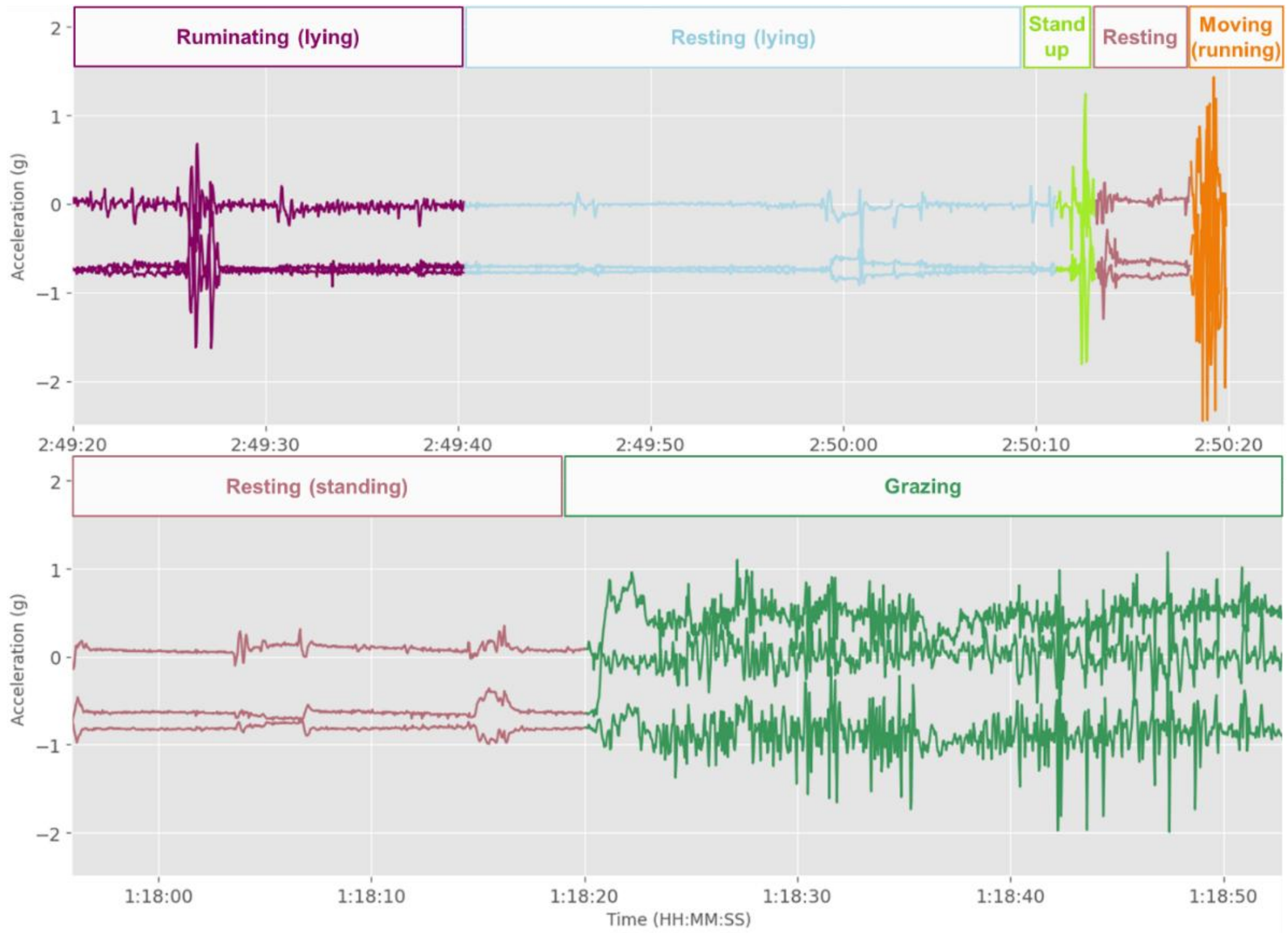

Figure 12. Two accelerometer sequences with the corresponding annotated behaviours for a single ewe [batch 1]. The three time-series in each sequence correspond to the X-axis, Y-axis and Z-axis.

for the different behaviours are displayed in Figure 12.

For batch 2, accelerometer data corresponding to each move for each ewe were identified using the start and end times of the move recorded with a smartphone clock. The corresponding accelerometer readings were displayed from 10 minutes before to 10 minutes after the move to carefully select the sequences corresponding to the moving pattern (see Figure 13). Accelerometer data within these sequences were thoroughly checked by a domain expert to remove any instances when ewes engaged in other activities, such as grazing along the track or standing still. These instances were identified using GoPro video and audio recordings collected during the move. Once inspected and validated, each moving session was merged with the corresponding accelerometer readings, and the behaviour "Moving" with the modifier "Moving along the track" (see Table 1 & Table 2) was added to the dataset, irrespective of the type of motion (walking, trotting, running). In total, more than 13 hours of moving-annotated accelerometer data from 82 ewes were obtained in batch 2. These moving sequences were finally merged with the annotated sequences from batch 1 using Python (V3.11.7), resulting in a combined dataset of 79 hours from 120 ewes.

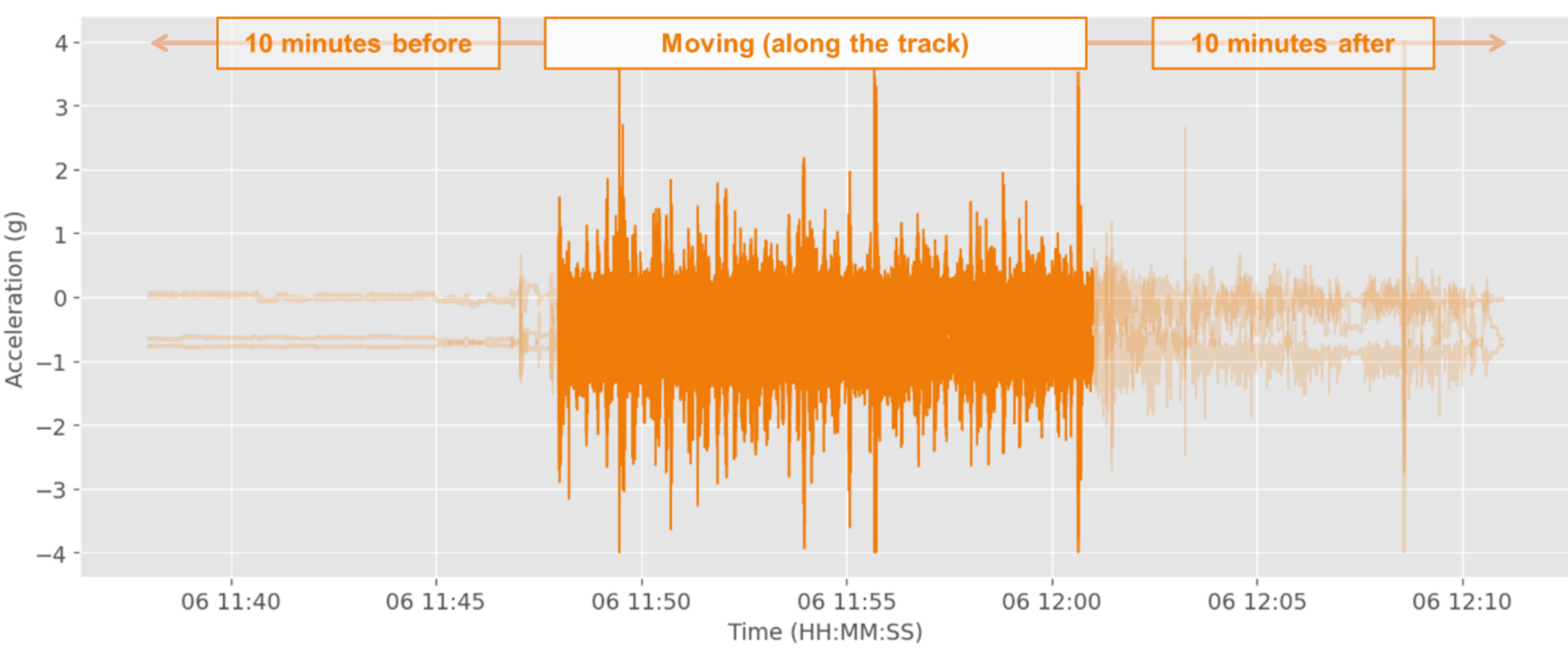

Figure 13. Accelerometer sequence recorded while one of the ewes was moving along the track [batch 2]. The three-time series correspond to the X-axis, Y-axis, and Z-axis; they are not shown for clarity.

# LIMITATIONS

Although considerable effort was made to include diverse conditions encountered in extensive systems, the animals used in this experiment were relatively homogeneous. Only female sheep of Romane breed, mostly born between 2023 and 2024 (*i.e.*, ewe lambs at the trial time), were equipped with the accelerometer collars. This specificity reflects the main objective of the EweAcT dataset: supporting the development of an AI model for monitoring the behaviour of ewe lambs from divergent genetic lines for social traits in extensive systems. Therefore, its applicability to adult sheep of other breeds remains uncertain and would require additional validation. A second limitation concerns data collected under hot conditions. Despite efforts to sample different periods, much of the Resting and Ruminating behaviours in the dataset were actually obtained when temperatures exceeded 20°C, with animals exhibited heat-related behaviours (e.g., moving frequently, visiting drinking troughs, flicking ears and heads). Therefore, collecting additional accelerometer data under cooler conditions for the Resting and Ruminating behaviour would help balance the dataset. Finally, behavioural annotation remains resource-intensive. Two observers spent several months annotating the videos. Therefore, we hope that this curated, ready-to-use dataset will speed up the development of robust and reliable AI models for monitoring sheep behaviour on pasture.

## ETHICS STATEMENT

All procedures involving animals were carried out in accordance with the ARRIVE guidelines. Ethical approval was obtained from the Ethics Committee in Science and Animal Health (No. 115) of the Veterinary School of Toulouse (Toulouse, France) under authorization ID SSA_2024_003V2. Only female sheep were used in the experiment, and all efforts were made to minimize stress and ensure animal welfare.

## CRediT AUTHOR STATEMENT

**Lucile Riaboff**: Methodology, Investigation, Data curation, Visualization, Original draft preparation, Writing, Supervision. **Ny Aina Andriamampandry**: Investigation, Data curation, Visualization, Original draft preparation. **Jean-François Bompa**: Conceptualization, Methodology, Investigation, Supervision. **Mathias Aletru**: Conceptualization, Methodology, Investigation. **Christian Durand**: Investigation, Supervision. **Sébastien Douls**: Investigation. **Gaëtan Bonnafe**: Investigation. **Morgane Costes-Thiré**: Investigation. **Guillaume Delosières**: Conceptualization, Methodology, **Jean-Marc Mongrelet**: Conceptualization, Software, Investigation. **Enzo Niro**: Conceptualization, Methodology. **Némuel Tadi**: Investigation. **Sara Parisot, Severine Deretz**: Resources. **Margot Lamarque**: Methodology, Conceptualization, Supervision. **Dominique Hazard**: Resources, Funding Acquisition, Writing- Reviewing and Editing, Supervision. **Emilie Cobo**: Conceptualization, Methodology, Investigation, Supervision.

## ACKNOWLEDGEMENTS

We would like to thank all the interns (Lizéa Gaubert, Eléonore Portes, Gabrielle Fourier and Hélène Largant) who assisted at the La Fage Experimental Unit during the 2024 trial, as well as Avelyne Villain and Nicolas Lesueur for their contributions to the data collection in 2025. This work was supported by state funding managed by the French National Research Agency (ANR) under the France 2030 program (PEPR Agroecology and Numeric), with funded projects WAIT4 and PATASEL (references ANR-22-PEAE-0008, ANR-22-PEAE-0013), as well as by the Genetics Animal division of the French National Research Institute for Agriculture, Food and Environment (INRAE).

## DECLARATION OF COMPETING INTERESTS

The authors declare that they have no known competing financial interests or personal relationships that could have appeared to influence the work reported in this paper.